\documentclass[preprint]{aastex}
\usepackage{spr-astr-addons}
\usepackage{url}

\RequirePackage{color}

\begin{document}
   \title{Foreground removal from CMB temperature maps using an MLP neural network}
\shorttitle{Foreground removal by a neural network}

\author{H.\,U.\,N{\o}rgaard-Nielsen}
\affil{National Space Institute,\\
 Technical University of Denmark\\
 Juliane Mariesvej 30, DK-2100 Copenhagen, Denmark\\
email: hunn@space.dtu.dk}
\and
\author{H.\,E.\,J{\o}rgensen}
\affil{The Niels Bohr Institute,\\
 Copenhagen University\\
Juliane Mariesvej 30, DK-2100 Copenhagen, Denmark}
\and

\shortauthor{N{\o}rgaard-Nielsen and J{\o}rgensen: Cleaning of CMB maps by a neural network}




  \begin{abstract}
   {One of the main obstacles in extracting the Cosmic Microwave Background (CMB) signal from observations in the mm-submm range is the foreground contamination by emission from galactic components: mainly synchrotron, free-free and thermal dust emission. Due to the statistical nature of the intrinsic CMB signal it is essential to minimize the systematic errors in the CMB temperature determinations.

   Following the available knowledge of the spectral behavior of the galactic foregrounds simple, power law-like spectra have been assumed. The feasibility of using a simple neural network for extracting the CMB temperature signal from the combined CMB and  foreground signals has been investigated. As a specific example, we have analysed simulated data, like that expected from the ESA Planck Surveyor mission.
   A simple multilayer perceptron neural network with 2 hidden layers can provide temperature estimates, over more than 80 percent of the sky, that are to a high degree uncorrelated with the foreground signals. A single network will be able to cover the dynamic range of the Planck noise level over the entire sky.}
   \end{abstract}



  \maketitle
%

\section{Introduction}

   The Cosmic Microwave Background (CMB) is by far the most important data set available for cosmological investigations. The angular power spectrum of its temperature and polarization anisotropies contain unique information about the basic cosmological parameters. Since the discovery of the CMB by Penzias \& Wilson (1965), tremendous efforts on the instrumental side have been made to improve sensitivity and angular resolution. The latest major experiment is the successful, ongoing, \emph{Wilkinson Microwave Anisotropy Probe} (WMAP Bennett et al.\,2003a) and the next  step forward in CMB measurements will no doubt be the ESA Planck mission.

    Due to the statistical nature of the CMB signal it is essential to subtract carefully all signals from non-cosmological sources, before the data can be used in a cosmological context. A clear demonstration of this problem has been given by Tegmark et al.\ (2000). Their model included frequency dependence, angular scale dependence for each physical component, and variations in the frequency dependence across the sky.  From these simulations they calculated angular power spectra and extracted the basic cosmological parameters. For the foregrounds they assumed power laws, and made 3 different sets of assumptions about the power law parameters: Optimistic (O), Middle-of-The-Road (MID), Pessimistic (PESS). With observational errors like those for Planck, they found that the accuracy of most cosmological parameters is degraded by a factor of about 2 for the MID model, and by a factor of about 5 for the PESS model. Therefore, in order to exploit fully the scientific capability of an experiment like Planck a procedure for reliable removal of all non-cosmological signals is mandatory.

   On angular scales larger than about 30' the non-CMB signal is dominated by diffuse emission from our own galaxy (De Zotti et al.\ 1999). Synchrotron (Haslam et al.\ 1982) and free-free (Haffner, Reynolds \& Tufte 1999; Finkbeiner 2003)  emissions dominate below about 60--80 GHz, while at higher frequencies thermal dust emission takes over. On smaller angular scales the foreground fluctuations are dominated by several populations of extragalactic sources, each with different spectral behaviour: radio sources, dusty galaxies and the Sunyaev-Zeldovich effect from clusters of galaxies.

   The CMB radiation has a virtually perfect black body spectrum (Mather 1999), while all known non-cosmological signals have a clear non-thermal frequency dependence. Therefore, it should be possible to split the observed microwave signal into its different components. This depends critically, of course, on the observational accuracy and the number of frequencies observed.

    A lot of different methods for component separation of CMB signals have been investigated, including:
    \begin{itemize}
     \item Maximum Entropy Method (MEM): Hobson et al.\ \cite{hobs98}, Stolyarov et al.\ \cite{stol02}, Barreiro et al.\ \cite{barr04}, Stolyarov et al.\ \cite{stol05}
     \item Internal Linear Combination method (ILC): e.g.\ Bennett et al.\ \cite{benn03b}
     \item Wiener Filtering: e.g.\ Tegmark \& Efstathiou \cite{tegm96}
     \item Independent Component Analysis (ICA) method: e.g.\ Maino et al.\ \cite{main02}
\end{itemize}
It is beyond the scope of this paper to give a review of all relevant component separation methods, but an excellent review can be found in  Delabrouille \& Cardoso \cite{dela07}.

    The statistical properties of the CMB temperature fluctuations on a sphere are normally expressed as a sum of spherical harmonics:
    \begin{equation}
    T(\theta,\varphi) = \sum_{l=0}^{\infty}\sum_{m=-l}^{l} a_{lm} Y_{lm}(\theta,\varphi),
    \end{equation}
    where $Y_{lm}(\theta,\phi)$ are the spherical harmonic functions. The simplest versions of the inflation paradigm predict that the CMB temperature fluctuations will be isotropically, randomly, Gaussianly distributed on a sphere. With this assumption, the statistical properties are then completely specified by the second order statistics, the angular power spectrum $C_{l}$,
    \begin{equation}
    C_{l}~=~\frac{1}{2l + 1}~\sum_{m=-l}|a_{lm}|^{2}.
    \end{equation}
   Therefore, a lot of effort has been put into constructing algorithms for foreground removal from CMB data, which assure minimum systematic residuals in the derived angular power spectrum. But, of course, an optimal method will also assure minimal systematic residuals in the CMB map itself.

   As demonstrated by Chiang \& Naselsky (2007), important information can be derived from the phases of the spherical harmonic coefficients. They find clear evidence for systematics in the distribution of phases at high latitudes in the WMAP 3y ILC map, and interpret them as due to residual foreground signals. For the Planck mission, one of the key scientific issues will be the search for primordial non-gaussianity in the CMB map. If clear evidence is not found, then tight upper limits on non-gaussianity will be important, emphasizing the need not only power for spectra with no systematic errors, but also maps.

   The main purpose of this paper is to investigate the feasibility of removing galactic foreground emission from CMB data by means of a neural network. As an example we will specifically examine simulated temperature data as expected from the ESA Planck CMB mission. See ESA-SCI(2005)1 for a detailed technical description of the payload and a discussion of the expected scientific capability.  Planck will provide maps of the microwave sky with unprecedented sensitivity and angular resolution. The 2 detector systems have been designed to cover a very broad frequency range in order to facilitate the removal of foregrounds. Therefore, it is important to analyse how well the different foreground signals can be removed by using only Planck data, and so-called `blind' extraction methods.

   For Planck, there is still uncertainty about how this component separation shall be performed in practice. Ideally, for each pixel on the sky the component separation procedure should provide an estimate of the temperature of the CMB, with an accidental error derived from the observational accuracy and with  negligible systematic errors (i.e.\ an error uncorrelated with any of the parameters of the different foreground signals).

    Several methods have been developed to estimate the CMB temperature without any assumptions about priors. Totally 'blind' methods like the 'Independent Component Analysis' method used by  FastICA (Maino et al.\ 2002) and SMICA (Delabrouille et al.\ 2003; Moudden et al.\ 2005) can, in principle, estimate both the CMB map and  the spatial and frequency dependence of the foreground sources, only relying on the CMB data itself. Similarly, in the 'Internal Linear Combination' (ILC) method used by the WMAP team (Bennett et al.\  2003b; Hinshaw et al.\ 2007), no assumptions are made \emph{a priori} about the foregrounds.
   Another approach was investigated by Brandt et al.\ (1994). They parameterized the spectra of the different foregrounds and then fitted them, sky pixel by sky pixel, with a non-linear least squares Levenberg-Marquardt algorithm (Levenberg 1944; Marquardt 1963). This approach has been further developed by Linden-V{\o}rnle \& N{\o}rgaard-Nielsen (1998) and Eriksen et al.\ (2006).

   In this paper we will elaborate on the ILC method by incorporating the non-linear features of neural networks. To set up these neural networks we parameterize the spectra of the different foregrounds in line with the Brandt et al.\ (1994) approach.

   This paper is organized in the following way:
   In Section 2, we outline the available information about spectral behaviour of the galactic foregrounds.
   We present the details of the modeling used in Sections 3 and 4.
   In Section 5, we briefly discuss the results found previously with the ILC method.
   Since neural networks are not commonly used in astrophysical literature, a brief introduction to the neural network concept is given in Section 6.
   We present the neural network used in Section 7 .
   In Section 8, we discuss our results in relation to similar previous investigations.

\section{Foreground modeling}
In the following, we identify the different non-CMB components to be included in our analysis, and give the details of the spectral models used.
\subsection{Synchrotron emission}
The galactic synchrotron emission originates from relativistic electrons spiraling in magnetic fields. In the galactic plane the magnetic field is ordered on large scales, with the field parallel to the spiral arms. Superimposed are small scale structures showing variations between the arm and inter-arm regions and with the local gas phase. The two components seem to have about the same magnitude.
At high latitudes, there are contributions from the galactic halo and specific structures e.g.\ the North Polar Spur. Variations in the spectral index come from variations in the electron energy spectrum, depending on the age and the origin of the electrons (e.g.\ supernova, diffuse shocks in the interstellar medium).
The synchrotron emission is traditionally modeled by a single power law
\begin{equation}
s_{s}(\nu)~=~A_{s}\left(\frac{\nu}{\nu_{0}}\right)^{\alpha_{s}},
\end{equation}
where $A_{s}$ is the synchrotron flux density at some reference frequency $\nu_{0}$ and $\alpha_{s}$ is the synchrotron spectral index. It is most likely that the spectral index will depend both on the frequency and the position on the sky, implying that at least two free parameters are needed to describe the synchrotron emission in a specific direction.

Brighter regions away from the galactic plane have typical values of $\alpha$ at 100 and 800 MHz of $-0.55$ and $-0.8$, respectively, Lawson at al. \cite{laws87}. At higher frequencies, the spectral index is expected to steepen by about 0.5 due to electron energy losses (Platania et al.\  \cite{plat98}). Banday et al.\ \cite{band03} derived a mean spectral index between 408 MHz and 19.2 GHz from the Cottenham \cite{cott87} survey, and between 31.5, 53 and 90 GHz from COBE-DMR data. The steep spectral index of $~ -1.1$, for galactic latitude $|b|>15^{0}$, is consistent with expectations. Bennett et al.\ \cite{benn03b} claim that the steepening occurs near the K-band (23 GHz). Eriksen et al.\ \cite{erik06} conclude that for $|b|>15^{0}$ the spectral index above 10 GHz is likely between $-0.7$ and $-1.2$ . As emphasized by Bennet et al.(1992) the synchrotron spectral index is dependent on the interstellar magnetic fields and is expected to steepen for higher frequencies. For example, a change of $B_{eff}$ from 0.1 to 5 $\mu$G gives a change in spectral index of 0.1 in the range 53--90 GHz.

\subsection{Free-free emission}
The Coulomb interaction between the free electrons and ions in the Milky Way results in thermal bremsstrahl-ung radiation, traditionally called free-free emission. Free-free emission is difficult to observe and simulate because at high latitudes it is not dominant at any frequency.

From the formulations given by Dickenson et al.\ \cite{dick03}, the free-free brightness temperature can be described with the following expression:
\begin{equation}
T_{ff,b} ~\propto~\nu^{-2}T_{e}^{-0.5}(\ln [0.04995\nu^{-1}]~+~1.5\ln T_{e}),
\end{equation}
where $T_{e}$ is the electron temperature.

Shaver et al.\ \cite{shav83} derived the electron temperature of HII regions at the galactocentric radius of the Sun as 7200 K $\pm$ 1200 K by means of radio recombination lines. Similar results have been found for a larger sample containing many weaker sources: Paladini et al.\ \cite{pala05}. At high latitude, the ionized gas will typically be within about 1 kpc of the Sun, and the electron temperature is should be in the range 7000--8000\,K, Dickenson et al.\ \cite{dick03}. It is possible that the diffuse emission at a given galactocentric distance may differ from the emission of high density HII regions in the galactic plane, as emphasized by Eriksen et al.\ \cite{erik06}.

From these constraints, it is expected that the effective spectral index in the frequency range relevant for Planck is $\alpha_{ff}=-0.14$. Between 10 and 100 GHz $\alpha$ values between $-0.1$ and $-0.2$ seem reasonable, and a steepening to $-0.3$ at hundreds of GHz is foreseen (Eriksen et al.\ \cite{erik06}).

Free-free emission is normally modeled by a simple power law
\begin{equation}
s_{ff}(\nu) ~=~A_{ff}\left(\frac{\nu}{\nu_{0}}\right)^{\alpha_{ff}}.
\end{equation}
Both Eriksen et al.\ \cite{erik06} and the WMAP team (Bennett et al.\ 2003, Hinshaw et al.\ \cite{hins07}) assume a constant free-free spectral index. Future high quality radio observations will probably show that this assumption is not valid, but it is still reasonable to expect that the spectral differences within the free-free galactic component will be smaller than for other foregrounds.

\subsection{Thermal dust emission}
Dust grains in the interstellar medium large enough to be in thermal equilibrium with the stellar radiation field will emit in the frequency range interesting for CMB research. From the all-sky observation of IRAS and COBE DIRBE it is known that this thermal emission peaks around 140$\mu m$.

Widely accepted models of the dust emission are extrapolations of the high-frequency IRAS, COBE DIRBE and FIRAS observations to CMB frequencies by Finkbeiner et al.\ (1999, hereafter FDS). These models use combinations of modified blackbody functions with different dust temperatures. These combined functions can approximate the integrated contributions from multiple components of dust, i.e.\ differences in grain properties, size, chemical composition and equilibrium temperature. The best-fit model of FDS model 8 assumes two main components with a total of 6 parameters. As emphasized by Eriksen et al.\ \cite{erik06} neither Planck nor any planned future CMB experiments have sufficient frequency resolution and sensitivity to constraint so many parameters for dust emission by themselves.

In their component separation analysis, Eriksen et al.\ \cite{erik06} calculate the dust emission from FDS model 8, but extract the dust component by using a much simpler function (FDS model 3), namely a power law combined with a slowly decreasing function of frequency of the order unity over the frequencies of interest (see Eq. 6).

\subsection{Anomalous dust emission}
By combining the COBE DMR maps with the DIRBE thermal dust emission map at 140$\mu$m, Kogut et al.\ \cite{kogu96} found evidence for an anomalous component with a rising spectrum from 53 to 31.5 GHz. The nature of this component has been widely discussed since. Hinshaw et al.\ \cite{hins07} give a detailed overview of the discussion so far. They conclude that the topic of the anomalous dust emission remains unsettled, probably until high quality diffuse measurements in the 5--15 GHz range are available for a significant part of the sky. Due to the uncertainty of its spectral behaviour and its detailed relation to the galactic dust component, we will not incorporate this component into the present investigation.

\section{Parameter ranges chosen for the neural network}

As emphasized by Eriksen et al.\ \cite{erik06}, neither Planck nor any of the planned future CMB missions will have sufficient power to constrain very detailed models of the galactic components, so simplified models will be needed. The situation may be improved by introducing additional external data to the fit. Since this paper will investigate the Planck data alone, this possibility is outside the scope of the paper.

Concerning the CMB temperature range, by examining the WMAP ILC map it follows that a range of $\pm 750\,\mu$K covers the dynamic range of the CMB thermodynamic temperatures well.

\begin{figure}[h]
\centering
\includegraphics[width=3.0 in]{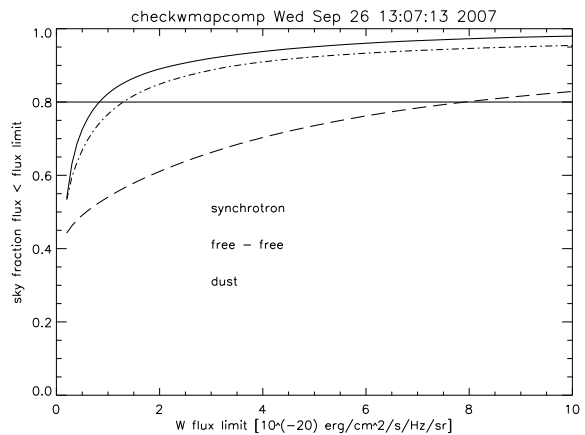}
\caption{Sky fraction covered by the different galactic components as a function of the WMAP W-band (93 GHz) flux. From top to bottom the three lines give the coverage of synchrotron emission, free-free emission and thermal dust.}
\label{figskyfrac}
\end{figure}

\subsection{Amplitude ranges at 100 GHZ}
We have normalized the component spectra at 100 GHz. The range of fluxes at this frequency  has been determined from the WMAP 3yr MEM maps for the three galactic components. The sky coverage as a function of the W-band (93 GHz) flux limit can be seen in Fig.~\ref{figskyfrac}. We have chosen a sky coverage close to the WMAP Kp0 mask, giving about 80 per cent coverage. The applied amplitude ranges are given in Table~\ref{tabpar}.
\begin{table}
\caption{Models parameters used for the neural network. The range of flux amplitudes at 100 GHz for the galactic components  are given in units of $ 10^{-20} erg/cm^{2}/s/Hz/sr$, while the unit of the CMB thermodynamic temperature is  $\mu$K.}              
\label{tabpar}      
\centering                          
\begin{tabular}{c c c}        
Component & Amp range  & Spec index range\\    
\hline                        
synchrotron & 0 : 1.0 & $-1.2$ : -0.6\\ 
free-free & 0 : 2.0 & $-0.2$ : $-0.1$\\
thermal dust& 0 : 8.0 & +2.75 : +3.75\\
CMB temp & $-750$:750\\
\hline                                   
\end{tabular}
\end{table}

\subsection{Ranges for the spectral indices}

As discussed above, it is expected that the synchrotron emission will not follow a simple power law with a constant spectral index, but the spectrum will steepen at higher frequencies. To take this into account in our simulations we have used a steepening of $-0.1$ at 44~GHz and $-0.20$ for the higher frequencies. The range of spectral indices (between 33 and 44 GHz) used in our simulations is given in Table~\ref{tabpar}. The variation in the spectral index is similar to the Tegmark et al.\ \cite{tegm00} PESS synchrotron model.

The variations in the spectral index of the free-free emission are expected to be small, and traditionally it is assumed to be constant (e.g.\ Eriksen et al.\ \cite{erik06}). In our simulations we have assumed a range between $-0.2$ and $-0.1$, similar to the range used in the PESS model by Tegmark et al.\ \cite{tegm00}.

To simplify the model of the thermal dust emission we have, following Eriksen et al.\ \cite{erik06}, used the single component  FDS model 3, with a temperature $T_{1} = 18.1 \rm{K}$. So we have
\begin{equation}
s_{d}(\nu)~\sim ~
\left(
    \frac{\exp \left( \frac{h\nu_{0,d}}{kT_{1}}\right) - 1}
    {\exp \left( \frac{h\nu}{kT_{1}}\right) - 1}
\right)
\left(
    \frac{\nu}{\nu_{0,d}}
\right)^
{\alpha_{d}+1}
 \end{equation}
 where $\alpha_{d}$ is the dust spectral index and
 $\nu_{0,d}$ = 3000\,GHz.

 \begin{figure}[h]
 \centering
\includegraphics[width=3.0in]{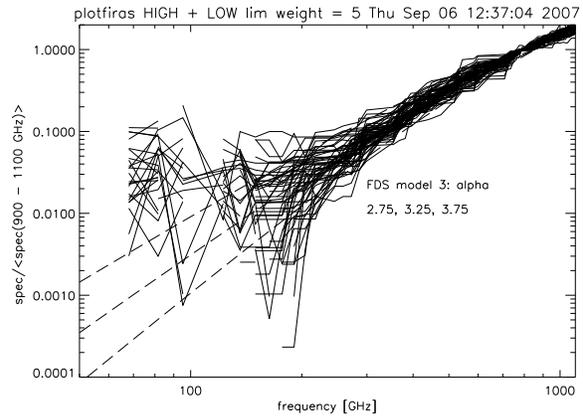}
\caption{Good quality FIRAS spectra (weight $\geq$ 5), median filtered (N = 10) and FDS model 3 with spectral indices 2.75, 3.25 and 3.75. The excesses seen at low frequencies for some of the spectra are due to synchrotron and/or free-free emission}
\label{figfiras}
\end{figure}

In Fig.~\ref{figfiras}, FIRAS spectra, taken from lambda at www.gsfc.nasa.gov, with weights $\geq$ 5 have been used, smoothed with a median filter  (n = 10), then normalised at 850 GHz. It is seen that the range of spectral indices given in Table ~\ref{tabpar} covers the dynamic range of the spectra well. The range is comparable to the PESS model of Tegmark et al.\ (2000).

\begin{figure}[h]
 \centering
\includegraphics[width=3.0in]{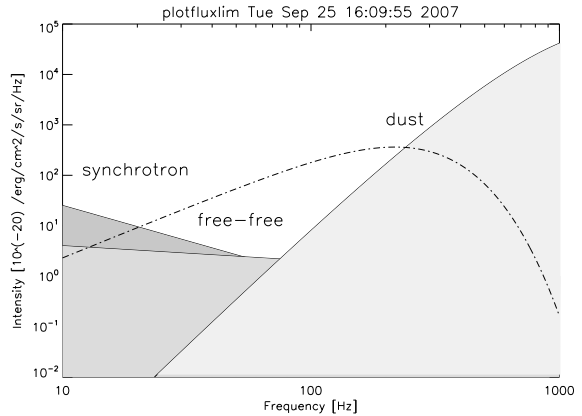}
\caption{The ranges of the galactic foregrounds covered by the simulations using the parameters from Table~\ref{tabpar}. The spectrum of a black body with $\delta T = 0.75\,\rm{mK}$ is shown as a dashed-dotted line}
\label{figfluxlim}
\end{figure}

The ranges of the galactic foreground models applied in the simulations are illustrated in Fig.~\ref{figfluxlim}.

\section{The ESA CMB mission: Planck}

The main scientific goal of the next ESA medium class mission, Planck, is to measure the CMB sky with unprecedented sensitivity and angular resolution. One of the main drivers of the design of the payload has been to assure proper removal of  non-CMB footprints from the maps. Planck contains 2 detector systems: the Low Frequency Instrument (LFI) based on HEMT technology (Principal Investigator N.\,Mandolesi), and the High Frequency Instrument (HFI) based on bolometers (Principal Investigator J.\,L.\,Puget). The reflector system is provided by ESA and a Danish consortium (Principal Investigator H.\,U.\,N{\o}rgaard-Nielsen). LFI covers 27--77 GHz, while HFI covers 67--1000 GHz, much wider than the CMB peak around 200 GHz. All LFI detectors are polarization-sensitive, while HFI has a number of polarization-sensitive bolometers for all frequencies less than 545 GHz. In this paper we examine the component separation problem for the Planck temperature measurements alone.
The expected sensitivity of the Planck detector systems is given in Table~\ref{tabsen}. Due to the arrangement of the Planck detectors in the focal plane, and the expected scanning strategy, it is reasonable to assume that the 1$\sigma$ values for the different frequencies will scale with the same factor, depending on the exposure time.
\begin{table*}
\caption{Summary of the Planck instrument characterization for a sky pixel with average exposure time, taken from the `Blue Book', ESA-SCI(2005)1. The 1$\sigma$ flux sensitivity is derived for a common beam with a 30 $\arcmin$ diameter (unit: $10^{-20} erg ~s^{-1}cm^{-2}Hz^{-1}sr^{-1}$)}              
\label{tabsen}      
\centering                          
\begin{tabular}{l c c c c c c c c c}        
 & LFI & LFI & LFI & HFI & HFI & HFI & HFI & HFI & HFI \\
\hline                        
 Center frequency  [GHz] & 30 & 44 & 70 & 100 & 143 & 217 & 353 & 545 & 857 \\    
Bandwidth ($\Delta\nu/\nu$)& 0.2 & 0.2 & 0.2 & 0.33 & 0.33 & 0.33& 0.33& 0.33 & 0.33\\
Angular resolution [arcmin] & 33& 24 & 14 & 10 & 7 & 5 & 5 & 5 & 5  \\ 
1$\sigma$ sensitivity & 0.22 & 0.38 & 0.90 & 0.58 & 0.61 & 1.19 & 2.25 & 4.34 & 5.22 \\
1$\sigma$ sensitivity [$\mu$K]& 6.76 & 6.71 & 6.77 & 2.43 & 1.61 & 2.46 & 7.59 & 76.09 & 3644.24 \\

\hline                                   
\end{tabular}
\end{table*}

\section{The Internal Linear Combination (ILC) method}

The ILC method was introduced by the WMAP team in their analysis of the 1 year data, Bennett et al.\ \cite{benn03b}. The basic assumption is that with a linear combination of the 5 WMAP frequency maps the foreground signals can be eliminated to a large extent and the cosmological signal extracted. With the rather complex non-linear functionalities of the foreground signals, this assumption will only make sense on rather small scales. The WMAP team divided the sky into 12 regions, of which 11 cover the galactic plane (see Hinshaw et al.\ \cite{hins07} Fig 8). Due to the basic limitations of the method, the WMAP team warned against using the ILC map for cosmological investigations.

Eriksen et al.\ \cite{erik04} have re-analysed the WMAP ILC map and improved the variance of the map significantly by introducing a Lagrange multiplier fitting algorithm. Using detailed Monte Carlo simulations, they investigated the limitations of the method and emphasized that great care needs to be taken both in its implementation and in understanding the influence of the residual foregrounds on the cosmological results.

\begin{figure}[h]
\centering
\includegraphics[width=3.0in]{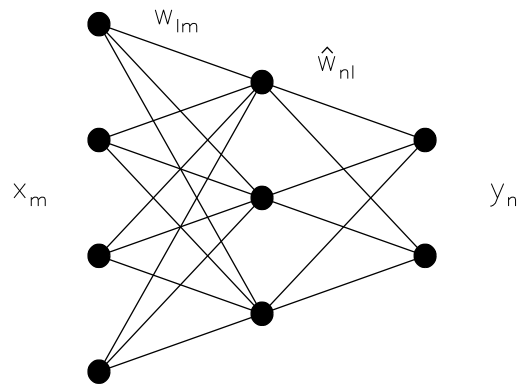}
\caption{A sketch of a \textbf{M}ulti\textbf{L}ayer \textbf{P}erceptron network with 1 hidden layer}
\label{fignnet}
\end{figure}

\section{Brief description of  the neural network concept}

    Neural networks and other 'machine learning' methods are not commonly used in astrophysical investigations. Recently, a few applications have focused on fast cosmological parameter estimations from CMB power spectra (Auld et al.\ \cite{auld07}, Fendt et al.\ \cite{fend07},  Habib et al.\ \cite{habi07}). Here we are investigating the feasibility of neural networks to obtain systematic-free CMB temperature estimates from noisy millimeter/submillimeter sky maps.

    Neural networks are analog computational systems whose structure is inspired by studies of the human brain. Many different architectures of neural network have been developed to tackle a variety of problems. An excellent introduction to neural networks can be found in Bishop \cite{bish95}. In the current paper we have only investigated one of the most simple and also most popular networks, namely the multilayer perceptron (MLP).

    An MLP consists of a network of units (called neurons) as illustrated in Fig.~\ref{fignnet}. Each unit is shown as a circle and the lines connecting them are known as weights. The network can be understood as an analytical mapping between a set of input variables $x_{m}~(m = 1,...,M)$  and a set of output variables $y_{n}~(n=1,...,N)$. The input variables are applied to the M input units on the left of the figure: M = 4 and N = 2 in the shown example. These variables are multiplied by a matrix of parameters $w_{lm}~(l = 1,...,L),~(m=1,...,M)$ corresponding to the first layer of links. Here L is the number of neurons in the middle (hidden) layer: L = 3 in the shown example. This results in a vector of inputs to the units in the hidden layer. Each component of this vector is then transformed by a non-linear function F, so we have
    \begin{equation}
    z_{l}~=~F \left( \sum_{m=1}^{M}~w_{lm}x_{m}~+\Theta_{l} \right) ~~(l=1,...,L), \label{eq1}
    \end{equation}
    where $\Theta_{l}$ is an offset (or threshold). We used the Neural Network Toolbox in the MATLAB software environment ($www.mathworks.com$) and have exploited the $tansig$ function as the non-linear function:
    \begin{equation}
    \rm{tansig}(x) ~= ~ \frac{2}{1~+~\exp (-2~x)}~-1, \label{eq2}
    \end{equation}
    It is seen that $tansig$ has an S-like shape, with values falling within the interval $[-1:1]$.
    From the hidden layer to the output units a linear transformation with weights $\widehat {w}_{nl}~(n=1,...,N;l=1,...,L)$ and offsets $\widehat{\Theta}_{n}$ are applied
    \begin{equation}
    y_{n}~=~\sum_{l=1}^{L}\widehat{w}_{nl}z_{l}~+~\widehat{\Theta}_{n}
        \quad\quad (n=1,...,N), \label{3}
    \end{equation}
    By combining Eqs.\ 1 and 2, it is seen that the entire network transforms the inputs $x_{m}$ to the output $y_{n}$ by the following analytical function
    \begin{equation}
    y_{n}(x_{1},...,x_{M})= \sum_{l=1}^{L}\widehat{w}_{nl}~F \left( \sum_{m=1}^{M}w_{lm}x_{m}+\Theta_{l}\right)+\widehat{\Theta}_{n}, \label{eq4}
    \end{equation}
    Clearly, such an MLP can easily be generalized to more than one hidden layer.

    Hornik, Stinchcombe \& White \cite{horn89} have shown that an MLP with a single hidden layer can approximate, with arbitrary accuracy, any non-linear multi-variate mapping, subject to only mild restrictions, provided the number of processing elements is large enough. Unfortunately, there is no general scheme for defining the optimal network, nor a guarantee that the network will converge within a reasonable timeframe.

    Given a set of P sample input and output vector pairs, $\{x_{m}^{p}~y_{n}^{p}\}~ p=1,...,P$, for a specific mapping, a technique known as error back propagation, can derive estimates of the parameters $w_{lm},~\Theta_{m}$ and   $\widehat{w}_{nl},~\widehat{\Theta}_{n}$, so that the network function (\ref{eq4}) will approximate the required mapping.
    The training algorithm minimizes the error function
    \begin{equation}
    E_{NET} ~=~\sum_{p=1}^{P}\sum_{n=1}^{N}[y_{n}(x^{p}) ~- ~y_{n}^{p}]^{2}, \label{eq5}
    \end{equation}
    The general concept for setting up a neural network is first to provide a test data set. This is traditionally split into  three data sets: one set used directly to train the network; a validation data set used in the iteration scheme, not directly in the training, but in the evaluation of the improvement of the network; and a test set which is only used at the end of the training to get an independent estimate of the accuracy of the derived network.

    Unfortunately, an MLP has a tendency, if allowed to iterate too many times, to reproduce accidental features in the test data, and therefore not provide optimal results when applied on completely independent data sets. Of course, you can follow how the iteration of the network is proceeding and stop it whenever it seems to fit the details of the test sample too closely. However, it is desirable to assure a good generalisation (also called regularization) of the network in an automated, unbiased, way.

    In MATLAB, an approach following the Bayesian framework, MacKay \cite{mack92}, has been implemented in the $trainbr$ procedure. The weights and offsets are assumed to be random variables with specified distributions. The regularisation parameters are related to the unknown variances associated with these distributions. In $trainbr$ the regulation part is combined with a training scheme following the Levenberg-Marquardt  non-linear least-squares method. An advantage of this procedure is that it provides a measure of how many parameters (weights and biases) are effectively used by the network, thus helping in setting up a reasonably sized network.

    Normally, neural networks are initiated by choosing completely random weights. Ngyen and Widrow (1990) have shown that by    choosing weights and bias values for each layer, so that the active  regions of its neurons are distributed approximately evenly over the input space, the network will be better able to form an approximation of any arbitrary function. The advantage of this method is that the network in general converges much faster. In MATLAB this method is implemented in $initnw$.

\begin{figure}[h]
\centering
\includegraphics[width=3.0in]{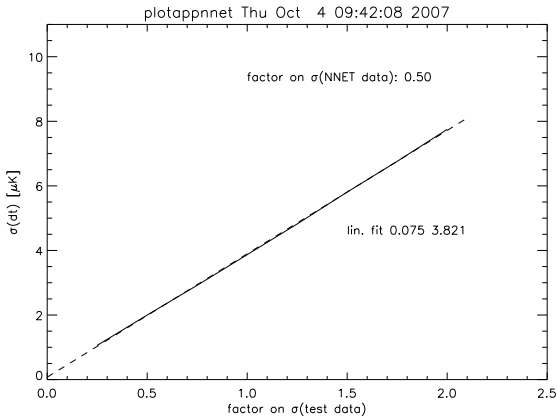}
\caption{The standard deviation of the $ \delta T$  determinations of a [12,3] MLP network (errors = 0.5 * Planck values) as a function of the error level of the test data (errors = f * Planck values, f = 0.25--2.0). Each test data set consists of 100,000 spectra, with parameters in the ranges specified in Table~\ref{tabpar}. It is seen that the relation closely follows a straight line (dashed) }
\label{fignnetobs}
\end{figure}

\begin{figure}[h]
\centering
\caption{Errors in the $\delta T$ determinations for a set of  100,000  independent simulated spectra ($f_{sen}$ = 1.5) run through the reference [12,3] network ($f_{sen}$ = 0.5) as function of the synchrotron constant $A_{s}$ and spectral index $\alpha_{s}$ . The correlation functions between the $ \delta T$  residuals and the 2 parameters are $-0.0125$ and 0.0427, respectively. A median filtering (N=100) of the residuals is given, and the extent of the 65 percent, 95 percent and 99 percent limits of the error distribution, derived in 10 sub-intervals, are shown as the solid horizontal lines, all displaced $-40 \mu\rm{K}$ for clarity. (Figure can be found at $ftp://ftp.spacecenter.dk pub/hunn/astro-fig6-9.zip$)}
\label{figres1}
\end{figure}

\begin{figure}[h]
\centering
\caption{Same residuals and network as Fig~\ref{figres1} but the residuals are plotted as a function of the free-free constant $A_{ff}$ and spectral index $\alpha_{ff}$. The correlation functions between the $\delta T$ residuals and the 2 parameters are $-0.0063$ and $-0.0065$. (Figure can be found at $ftp://ftp.spacecenter.dk/pub/hunn/astro-fig6-9.zip$)}
\label{figres2}
\end{figure}

\begin{figure}[h]
\centering
\caption{Same residuals and network as Fig~\ref{figres1} but the residuals are plotted as a function of the thermal dust constant $A_{d}$ and spectral index $\alpha_{d}$. The correlation functions between the $\delta T$ residuals and the 2 parameters are 0.0025 and 0.0158. (Figure can be found at $ftp://ftp.spacecenter.dk/pub/hunn/astro-fig6-9.zip$)}
\label{figres3}
\end{figure}

\begin{figure}[h]
\centering
\caption{Same residuals and network as Fig~\ref{figres1} but the residuals are plotted as a function of the input CMB temperature $\delta T$. The correlation function is 0.0003. (Figure can be found at $ftp://ftp.spacecenter.dk/pub/hunn/astro-fig6-9.zip$)}
\label{figres4}
\end{figure}

\section{\textbf{The applied neural network}}

Since Planck is scanning the sky in a rather inhomogeneous way,
it is expected that the observational errors will vary across the sky. Bernard et al.\ \cite{bern02} have simulated the exposure times for each sky pixel ($10\arcmin \times 10\arcmin$) for different assumptions about the scanning strategy. They found for all scanning models considered by the Planck Science Team  that the exposure times vary by only a factor of about 10 over the sky. Due to the design of the Planck focal plane it is reasonable to assume that sensitivities in the different frequency bands will vary with the same factor.
In Table~\ref{tabsen} we have given the expected observation sensitivity per frequency band for an average 30$\arcmin$ sky pixel. So it is expected that these sensitivities will vary by the same factor, $f_{sen}$, in the range 0.6--1.6 across the sky.

If symmetrical Gaussian beams and white noise are assumed, e.g.\ no $1/f$ noise contribution, the noise of the individual sky pixels is independent, and it is possible to treat each sky pixel separately.
These assumptions were also made in the Planck Working Group 2 Component Separation Challenge, see below.

Under these conditions, and assuming that the three foreground components can be treated as discussed in Section 2, with the parameter ranges specified in Table~\ref{tabpar}, it is straight forward to set up a suitable neural network.

In this paper only simple MLP networks have been investigated to examine their ability to remove foreground signals from CMB data.

We have set up networks to establish an algorithm for deriving the CMB temperature anisotropies from simulated spectra, using all 7 foreground parameters and including noise, covering the 9 Planck frequencies
 such that
$[obs_{i}, i = 1, 9]~\rightarrow~\delta T$.

To assure the feasibility of the network, it is, of course, important to assure that the expected parameter ranges are covered, (see Table~\ref{tabpar}), and that the different combinations of parameters are covered to a reasonable extent. Experience has shown that a training data set of about 10,000 spectra is sufficient.

To set up a neural network, the number of hidden layers, the number of neurons for each layer, the scale factor on the observational errors, $f_{sen}$, and the number of data in the training set, $N_{TRAIN}$, needs to be specified. For a running validation of the development of the network iteration process, a data set only for this purpose is also calculated with $N_{VAL}$ spectra, normally 25 percent of $N_{TRAIN}$.

The final network is then tested with a data set with $N_{TEST}$ spectra ($f_{sen}$ could be different than that for the network) derived completely independently of the 2 data sets used to train the network.

With these parameters fixed, the following procedure is used:
\begin{enumerate}
\item Draw 7 parameter values, uniformly distributed within the parameter ranges given in Table~\ref{tabpar}.
\item Calculate the combined fluxes of the 4 foreground components at the 9 Planck frequencies.
\item For each frequency add a Gaussian randomly distributed number, multiplied by 1$\sigma$ values given in Table~\ref{tabsen} and the assumed scaling factor $f_{sen}$.
\item Repeat 1--3 until the desired number, $N_{TRAIN}$, of spectra in the training set have been obtained.
    \item Repeat 1--3 until the desired number, $N_{VAL}$, of spectra in the validation set have been obtained.
\item Train the neural network to find the weights describing the mapping between the input spectra and the true temperature anisotropy, known for the simulated input.
\item Obtain independent test samples of spectra by repeating steps 1--3  $N_{TEST}$ times with different $f_{sen}$.
\item Run the $N_{\rm{TEST}}$ spectra through the network to get an independent estimate of the reliability of the network.
\end{enumerate}

The basic neural network used in this paper is an MLP with 2 hidden layers, with 12 and 3 neurons, respectively (referred to as a [12,3] network). The training set contained 10,000  spectra and the sensitivity scaling factor, $f_{sen}$, was 0.5.
To set up this kind of network, the requirements on computer power are quite modest. The training of the reference network took a total of about 15 CPU min.\ on a Sun Fire V40z (8 AMD Optron @ 2.2 GHz, 16 GB RAM).

To represent the range of sensitivities expected for Planck we have run several test data sets through the network, each with 100,000 spectra and with $f_{sen}$ varying from 0.25 to 2.0, covering the sensitivity range relevant for Planck.
Fig.\ \ref{fignnetobs} shows the rms of the CMB temperature estimates as a function of $f_{sen}$. It is seen that the relation follows to high accuracy a simple scaling over the full range of applied errors.

To demonstrate the level of systematic errors in the temperature estimates, Figs.~\ref{figres1}, \ref{figres2}, \ref{figres3} and \ref{figres4} show the results obtained by the reference neural network and the test data set with $f_{sen}~=~ 1.5$.
The figures show the difference between input CMB temperature and the temperature derived by the network plotted as a function of the 7 basic parameters in our sky model. It is seen that the distribution of $ \delta T- \delta T (\rm{NNET})$ is very close to Gaussian (skewness = 0.0133, kurtosis = 0.0129). In each figure the correlation
between the $\delta T$ residuals and the respective input parameter is also given:
$\rm{corr}(x,y) = <(x-<x>)(y-<y>)>/(\sigma_{x}\sigma_{y})$.

In the figures, the median (N = 100) filtered, reordered, residuals are plotted together with the extent of the 65 percent, 95 percent and 99 percent limits of the error distribution, obtained for 10 subintervals, shown as the horizontal lines, all shifted $-40 \mu$K for clarity.

It can be seen that errors of the CMB temperatures derived by the network are to a high degree uncorrelated with any of the basic input parameters. It can also be seen that the scatter of the $ \delta T$  determinations is independent of the amplitudes of the foreground components. This is in strong contrast to direct fitting, e.g.\ Levenberg-Marquardt, where the errors increase rapidly when the input amplitudes get close to zero.

\section{Discussion}

Brandt et al.\ (1994) simulated $10\degr\times 10\degr$ patches on the sky  with an angular resolution of $1\degr$ for various choices of observing frequencies. They assumed simple power laws for the synchrotron and free-free emissions, while the dust emission was assumed to be a combination of two black body spectra, ($T_{1} = 20.4 \rm{K}$ and $T_{2} = 4.77 \rm{K}$), with an emissivity power law having a  variable index and a variable scaling factor. They fitted the simulated spectra for each sky pixel by means of a Levenberg-Marquardt algorithm. They emphasized that this direct fitting is unstable due to the fact that the number of parameters is only slightly smaller than the number of observed frequencies, which is actually the case for all CMB experiments.

It is especially difficult to fit the synchrotron and free-free emissions simultaneously. This problem is normally smoothed over by combining the two components into a single power law in the fitting routine (Brandt et al.\ 1994; Linden-V{\o}rnle \& N{\o}rgaard-Nielsen 1998). Clearly, with this simplification systematic errors will be introduced at some level in the CMB temperature estimates. One way to handle this problem is to assume a constant spectral index for the free-free emission e.g.\ Eriksen et al.\ (2006).

Eriksen et al.\ (2006) elaborated further on the Brandt et al.\ approach. They assumed a simple power law for the synchrotron emission and a power law with a constant spectral index for the free-free emission. The thermal dust emission was assumed to follow  FDS model 8, but in the fitting routine they applied the much simpler FDS model 3. They set up a Markov Chain Monte Carlo (MCMC) algorithm to fit the spectrum of each individual sky pixel. They estimate that the fitting for each pixel will take about 100 seconds, implying that it is unfeasible to fit the spectra of all 50 million Planck sky pixels. Therefore, Eriksen et al.\ relaxed the angular resolution of the data to be fitted with the full MCMC algorithm. In their simulations they smoothed the original simulated maps (resolution $\sim$ 1 deg) with a Gaussian (FWHM 6 deg). These low-resolution data were then run through the MCMC algorithm. For the high resolution data they used the low-resolution non-linear parameters (power law spectral indices) and fitted only the linear parameters (the scaling factors).

Eriksen et al.\ give details of their MCMC fit only for a single pixel, in the galactic plane (l = 58\degr~and b = 0\degr). They emphasize that the parameters found for this pixel are clearly correlated (see their Figs. 3 and 4). The MCMC algorithm rejected the fit for this pixel at a high confidence level, mainly because the spectrum is fitted by an FDS model 3 and not an FDS model 8 spectrum like the input spectrum.
Eriksen et al.\ give the residual maps for the CMB temperature and for each of the component parameters, both for the low and high resolution data.
The problem with FDS model 3, versus model 8, is clearly evident at high latitudes in their goodness-of-fit plot (Eriksen et al.\ Fig. 6). They fixed the problem at high latitudes merely by introducing, by hand, a constant dust spectral index.
Eriksen et al.\ found that the distribution of CMB temperature residuals fitted reasonably well to the rms errors estimated by the MCMC algorithm.

Eriksen et al.\ do not provide information about the correlation between the errors in their CMB temperature estimates and the basic paramaters of their MCMC fitting scheme.

Hansen et al.\ (2006) have suggested that instead of removing the galactic foregrounds by means of a linear combination of observations obtained at different frequencies (i.e.\ ILC), a linear combination of differences between different frequencies should be used. Hansen et al.\ assume that the frequency spectra of the foreground components are independent of position on the sky. By analysing local regions separately this assumption can be relaxed. As emphasized by Hansen et al., this method increases the noise level in the data, implying that it is mainly useful for studies at larger scales where noise is not dominant.

The blind methods for component separation of CMB data are, of course, not totally blind. These methods model the total sky signal as a linear mixture of a few independent emission processes. The observation of the sky with a detector d is then a noisy linear mixture of $N_{c}$ components:

\begin{equation}
y_{d}(\theta,\phi) ~=~\sum_{j=1}^{N_{c}}~A_{dj}~s_{j}(\theta,\phi)~+~n_{d}(\theta,\phi)
\end{equation}
where $s_{j}$ is the emission template for the jth astrophysical process. The coefficients $A_{dj}$ reflect the emission laws and detector properties, while $n_{d}$ accounts for the noise. As emphasized by Delabrouille et al.\ (2003) these methods have 3 basic assumptions:

1) The mixing matrix, $A_{dj}$, is position-independent.

2) The components are statistically independent.

3) Noise is uncorrelated between the detectors.

In the real world it will be difficult to fulfill these requirements. The spectral indices and relative contributions of the different galactic components are known to change across the sky (e.g.\ the coefficients for the ILC fit of the WMAP 3 yr data, Hinshaw et al.\ 2007). With the improved sensitivity of Planck, it is reasonable to assume that even larger differences will be found. A way out of this problem is to isolate areas of the sky where assumption 1 is reasonable (Delabrouille et al.\ 2003).

Since all 3 galactic components are strongly concentrated towards the galactic plane, they are to some extent correlated. At higher latitudes the situation is unclear, but a detailed analysis of the correlations between the components will be one of the scientific tasks for Planck, so some interesting science could be missed by keeping assumption 2.

With the sophisticated and complex nature of Planck detector systems, combined with the weakness of the CMB anistropy signal, it will not be a surprise if the final sensitivity of the 9 frequency maps will first be achieved after very detailed investigation of all possible systematic errors (e.g.\ residual 1/f noise and temperature variations). It will also probably be too optimistic to expect that all the systematic errors will have been removed from the data.

Due to the importance of establishing a reliable procedure for removal for obtaining a foreground - free CMB map, Planck Working Group 2 has produced detailed maps with reasonable observational errors called the Planck Sky Model (PSM). As emphasized above, these maps are produced assuming symmetrical Gaussian beams and white observational noise, as in our neural networks. The maps have been run through 8 different separation algorithms and the preliminary results have been discussed by Leach et al.\ \cite{leac08}. As seen in their Fig.\ 5, on a colour scale of $\pm 30 \mu \rm{K}$, small systematic errors are present in the residual maps for all 8 algorithms. In a forthcoming paper we will analyse these PSM data, both for the simulated Planck and the WMAP maps by means of dedicated neural networks.

\section{Conclusions}
In this paper we investigate mainly the concept of using simple neural networks (MLPs) to remove the galactic foregrounds from the CMB signal, making some simplifying assumptions: only 3 galactic components are taken into account (synchrotron, free-free and thermal dust emission); the spectra of the foregrounds follow simple modifications to power laws; the observational noise is white (Gaussian); Gaussian beams are assumed. Under these assumption it has been demonstrated that for more than 80 per cent of the sky a neural network can provide reliable estimates of the CMB temperature with errors, which to a high degree are uncorrelated to the basic parameters of the sky model used.

The spectra of the foregrounds in the Planck frequency range, close to the galactic plane, are at present known to be complex, but little is known about the complexity at high latitudes. Therefore, in the case of Planck the details of the modeling of the foregrounds will first be established when the data has been carefully reduced and analysed. This will, no doubt, result in the need for including more galactic components (e.g.\ spinning dust) and, most likely, a revision of the spectral behavior of the components will be required. From the available knowledge of the Planck detectors, it is clear that the noise will not be white, and it will be a challenging task to investigate how to minimize the influence of non-white noise features.
It could well be that more sophisticated neural networks than MLP's will be needed in the Planck pipeline.

To further test the capabilities of neural networks to clean CMB temperature maps, we will, in a forthcoming paper, analyse the Planck Sky Model simulations of both Planck and WMAP data, as well as some real data, the recently released WMAP 5yr data.

\acknowledgments
We would like to thank Drs. G.\,De Zotti, J.\,Delabrouille and A.\,Hornstrup for constructive comments, and Dr. C.\,A.\,Oxborrow for significant improvements to the language and typesetting of this paper.

\end{document}